\documentclass[aps,prb,twocolumn,showpacs,floatfix,superscriptaddress,amssymb,amsmath]{revtex4}
\usepackage{graphicx}
\usepackage{dcolumn}
\usepackage{bm}
\usepackage{color}
\newcommand{\be}{\begin{equation}}
\newcommand{\ee}{\end{equation}}
\newcommand{\bea}{\begin{eqnarray}}
\newcommand{\eea}{\end{eqnarray}}
\newcommand{\p}{\partial}

\newcommand{\re}{\mbox{e}}
\newcommand{\ba}{\begin{array}}
\newcommand{\ea}{\end{array}}
\def\nn{\nonumber\\}

\def\up{\uparrow}
\def\down{\downarrow}

\begin{document}
  
  \title{Graphene zigzag ribbons, square lattice models and quantum spin chains }
  
  \author{Mahdi Zarea} 
  \author{Nancy Sandler} 
  \affiliation{Department of Physics and Astronomy, Nanoscale
    and Quantum Phenomena Institute, and Condensed Matter and 
Surface Science Program,\\Ohio University, Athens, Ohio
    45701-2979}
  
  \date{\today}
  
  \begin{abstract}

We present an extended study of finite width zigzag graphene ribbons (ZGRs) based on a tight-binding model with hard-wall boundary conditions. We provide an exact analytic solution that clarifies the origin of the predicted width-dependence on the conductance through junctions of ribbons with different widths. An analysis of the obtained solutions suggests a new description of ZGRs in terms of coupled chains. We pursue these ideas further by introducing a mapping between the ZGR model and the Hamiltonian for $N$ coupled quantum chains as described in terms of $2N$ Majorana fermions. The proposed mapping preserves the dependence of ribbon properties on its width thus rendering metallic ribbons for $N$ odd and zero-gap semiconductor ribbons for $N$ even. Furthermore, it reveals a close connection between the low-energy properties of the ZGR model and a continuous family of square lattice model Hamiltonians with similar width-dependent properties that includes the $\pi$-flux and the trivial square lattice models. As a further extension, we show that this new description makes it possible to identify various aspects of the physics of graphene ribbons with those predicted by models of quantum spin chains ($QSC$s).

 \end{abstract}%
 \pacs{75.10 Pq, , 3.20 At, 73.63.-b,71.30.+h  ,73.43.f} 
  
  \maketitle
Graphene has attracted a lot of interest since its successful synthesis in 2004\cite{Gaim}. The crystalline structure of the material is given by a triangular Bravais lattice with a two-atom 
basis which provides a spinor character to wave-functions. Calculations of the band structure reveal that
conduction and valence bands coincide at six corners of the Brillouin zone (the so-called Dirac points). The linear spectrum around these points makes possible to describe the low energy physics by
Dirac-type equations. Therefore graphene shows many unique properties not observed in other materials. Among them we should cite the unusual quantum Hall effect and the  high conductance of
charge carriers through potential barriers, a property  which is attributed to the Klein paradox \cite{Katsnelson}.

From the point of view of potential applications, a high mobility and stiffness makes graphene
an excellent candidate for future electronic devices. Interestingly enough, confined graphene samples (such as ribbons and quantum dots) reveal even more peculiar properties that strongly depend on edge terminations\cite{Nakada}. Thus, a cutting through armchair edges, renders ribbons that can be metallic or semiconducting, depending on the ribbon width, while $ZGR$s have localized states near the edges with
an almost flat band between any two inequivalent Dirac points \cite{Fujita}.  
Furthermore, numerical\cite{Li} and theoretical\cite{Beenakker3} work predicted a conductance of $ZGR$s strongly
dependent on the ribbon's width. These works, focused on transport properties through junctions, concluded that junctions between ribbons with even and odd number of zigzag lines showed a valley-valve effect.

In this work we show that this even-odd behavior has a simple explanation based 
on the presence or absence of zero modes in the spectrum of a finite width ribbon. This realization
leads naturally to a description of ZGRs in terms of coupled chain models that can be extended (in
some cases) to models of quantum spin chains (QSCs). This dual description suggests that $ZGR$s can be viewed as a material realization of many different theoretical models of QSCs. At the same time, the well-known physics of various $QSC$s models can shed light on various properties of graphene ribbons. 

\section{Model for zigzag graphene ribbons}
\label{sec1}

\begin{figure}[!]
  \includegraphics[width=.5\textwidth]{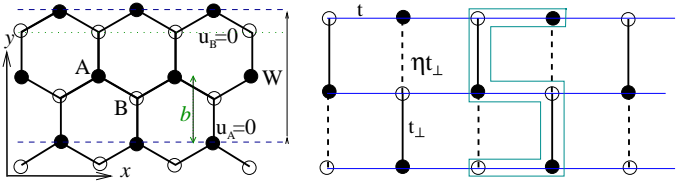}
  \caption{The boundary conditions for $ZGR$ implies 
that $u_A=0$ on the lower edge and $u_B=0$ on the upper edge 
(green dotted line). All atoms alone each zigzag line are labeled by the same
$y$-coordinate. The  right panel shows a generalized lattice model with interchain hoping
term $t_{\perp}$ and extra hoping term $\eta t_{\perp}$.
 In graphene $\eta=0$ and $t_{\perp}=t$. 
}
  \label{ZGR}
\end{figure}

We first review some results obtained with a tight-binding model for $ZGR$s with hard-wall boundary conditions.
In the absence of magnetic fields and spin-orbit interactions, the Hamiltonian is spin-independent and thus it is enough to consider spin up (or down) electrons only. The Hamiltonian and the spinor wave-vector in momentum space are given by:
\be
H  =  \left( \begin{array}{cccc}
  0 & \varphi \\
  \bar\varphi & 0 
\end{array}\right)
~~~~%
\Psi_{\pm}=C\left( \begin{array}{l}
 u_A= \re^{i\alpha/2}  \\
  u_B=\pm \re^{-i\alpha/2}\end{array}\right)\re^{ik_xx+ik_yy}
\ee
with $\varphi(k_x,k_y) =  t_{\perp}\re^{ik_yb}+2t\cos{\frac{k_xa}{2}}$ and
$\bar\varphi(k_x,k_y)=\varphi(k_x,-k_y)$. Here $k_x,k_y$ are measured from 
the center of the Brillouin zone (the $\Gamma$ point) and $b=a\sqrt{3}/2$. 
The angle $\alpha$ is defined by $\phi(k_x,k_y)=\sqrt{\varphi\bar\varphi}\re^{i\alpha}$.  
By an appropriate gauge transformation, all atoms along each zigzag line can be labeled by the same coordinate $y$. This is equivalent to work with the deformed lattice shown in Fig.\ref{ZGR}
Notice that in graphene the inter- and intra- chain couplings are equal ($t_{\perp}=t$). However for reasons to be clear
later, we introduce these couplings as independent variables. 

Standard hard-wall boundary conditions for $ZGR$s are given by
 $u_A(y=-W/2)=0$ and $u_B(y=W/2)=0$ along the lower and upper 
edge respectively\cite{Fujita,Nakada,Brey,Hikihara}. It is useful to introduce the
variable $N$ that counts the number of zigzag lines or chains inside the ribbon as $N=W/b-1$.
Note that due to translational symmetry along the $x-$ direction, $k_{x}$ is a good quantum number, and for a given $k_x$ the wave function is given by the 
linear combination of two degenerate states at $\pm k_y$
 \bea
\chi(k_x)_{\pm}&=& \left( \begin{array}{c} 
  \sin( k_yy+\alpha/2-n\pi/2)  \\
  \sin(k_yy-\alpha/2-n\pi/2)\end{array}\right)\label{wavegen}
\eea
in which  $k_y$ satisfies
\be
\alpha-k_yW=n\pi.
\label{zgr-cnd}
\ee
where $n$ is an integer. The condition imposed by Eq.~(\ref{zgr-cnd}) can also be written as $\varphi/\bar\varphi=\re^{2ik_yW}$. This also implies that
\be
\varphi=\re^{ik_yW}\sin k_yb/\sin k_yW \label{use}.
\ee

One peculiar feature of $ZGR$ is the presence of bands with complex wave number $k_{y}$  (and correspondingly complex $\alpha$) between two in-equivalent Dirac points
$K=(2\pi/3,0)$ and $K'=(4\pi/3,0)$. Previous works in the limit of semi-infinite ribbons have shown that these bands
are flat and correspond to highly localized states at the edge of the $ZGR$\cite{Fujita,Nakada,Hikihara,Brey}. These results can also be easily obtained from the definition of $\varphi$ and Eq.(\ref{use})  which supports a solution with  $\re^{ik_yb}\approx-2(t/t_{\perp})\cos(k_xa/2)$ with zero energy.

To get further insight into the properties of finite width ribbons we derive below the expressions for  wavefunctions and  energies of these quasi-degenerate bands.

First we consider the case where  $k_x>\pi$ that renders $\cos(k_{x}a/2) < 0$. For a zero energy solution to exist ($\varphi=0$), it is necessary that $\re^{ik_y} > 0$. This implies that $k_{y}$ should be imaginary and we can write $k_y=iq$ where $q$ is a positive real number. By returning to Eq.~(\ref{use}) we obtain
$\re^{i\alpha} > 0$ which implies that $\alpha$ is imaginary. Therefore  the condition set in Eq.~(\ref{zgr-cnd}) can be satisfied by setting $n=0$ and the wave-functions for the edge states are given by
\bea
\chi_{\pm}(k_xa>\pi))&=&C\left( \begin{array}{c} 
  \sinh q(y+W/2)  \\
  \pm\sinh q(y-W/2)\end{array}\right). 
  \label{edge_p}
\eea

Now consider the case for $k_x<\pi$ for which $\cos(k_{x}a/2) > 0$. The zero energy condition now requires that  $\re^{ik_yb}\approx-2\cos(k_xa/2)t/t_{\perp} < 0$ and thus $k_{y}$ must be a complex number: $k_y=\pi+iq$ where $q$ is a positive real number. 
In this case Eq.~(\ref{zgr-cnd}) can be satisfied by $n=N=W/b-1$ which results in 
\bea
\chi_{\pm}(k_xa<\pi)=C\left( \begin{array}{c} 
  \sinh [(q+i\pi)(y+W/2) \\
  \pm\sinh [(q+i\pi)(y-W/2)-iN\pi]\end{array}\right)
  \label{edge_m2}
\eea

The energy dispersion  for all bands  is given by $E=\pm t_{\perp}\sin(k_yb)/\sin(k_yW)$. In particular,
the edge state dispersion near $k_xa=\pi$ can be approximated as: 
\be
E\approx  \pm t_{\perp}(tk/t_{\perp})^N 
\label{EEd}
\ee
where $k$ it the dimensionless momentum measured from  the center of the band $k_{x}a=\pi$ (located
between two inequivalent Dirac points).

Eq.~(\ref{EEd}) reveals an important feature of $ZGR$: ribbons with an odd number of chains possess different transport properties than those with an even number of chains\cite{Li,Beenakker2}. 
This feature is reflected in the phase factor of $N\pi$ that appears in Eq.~(\ref{edge_m2})
which changes the relative sign of the spinor components in the wavefunction. A clear picture of this feature emerges when considering wavefunctions at three different $y-$points (one at the center of the ribbon $y=0$ and two points on opposite edges $y = \pm W/2$)  and finding their asymptotic form when approaching $k_xa=\pi$. 

The argument proceeds as follows: first, evaluate the wavefunctions at $k_xa>\pi$ and obtain their limiting expression as $k_xa\to \pi$. 

 At $y=0$ as $q\to \infty$  the spinor is:
\bea
\chi_{\pm}(k_xa>\pi)&\approx&C\left( \begin{array}{c} 
  \re^{qW/2}  \\
  \mp\re^{ qW/2}\end{array}\right)
\approx C\re^{qW/2}\left( \begin{array}{c} 
  1  \\
  \mp 1\end{array}\right)\label{x0p}
\eea

$C\sim \re^{-qW}$ and the wave function inside the ribbon goes to zero
at $k_xa=\pi$. Notice the relative sign of $u_A/u_B$.

At the left edge $y=-W/2$ as $q\to \infty$  the expression reduces to:
\bea
\chi^>_{\pm}&\approx&C\left( \begin{array}{c} 
  0  \\
  \mp\re^{ qW}\end{array}\right)
\approx C\re^{qW}\left( \begin{array}{c} 
  0  \\
  \mp 1\end{array}\right)\label{xlp}
\eea
and finally, at the right edge $y=W/2$, as $q\to \infty$  we obtain
\bea
\chi^>_{\pm}&\approx&C\left( \begin{array}{c} 
  \re^{qW}  \\
  0\end{array}\right)
\approx C\re^{qW}\left( \begin{array}{c} 
  1  \\
  0\end{array}\right)\label{xrp}
\eea

Note that for each band (conduction or valence) the signs of the spinor components
are consistent when we start from the left (Eq.~(\ref{xlp})) go through the middle (Eq.~(\ref{x0p}))
and then to the right (Eq.~(\ref{xrp})).
 
Now let us approach $\pi$ from the left, i.e., with $k_xa<\pi$: 

At $y=0$ as $q\to \infty$ the expression for the spinor is:

\bea
\chi_{\pm}(k_xa<\pi)&\approx&C\left( \begin{array}{c} 
  \re^{qW/2}\re^{i\pi W/2}  \\
  \pm\re^{ qW/2}\re^{-i\pi W/2}\end{array}\right)\\
&\approx& C\re^{qW/2}\left( \begin{array}{c} 
  \re^{i\pi W/2}   \\
  \pm \re^{-i\pi W/2} \end{array}\right)
  \label{x0m}
\eea

At the left edge $y=-W/2$ and as $q\to \infty$
\bea
\chi^<_{\pm}&\approx&C\left( \begin{array}{c} 
  0  \\
  \pm\re^{ qW}\end{array}\right)
\approx C\re^{qW}\left( \begin{array}{c} 
  0  \\
  \pm 1\end{array}\right)\label{xlm}
\eea
 Similarly, at the right edge $y=W/2$ and as $q\to \infty$ 
\bea
\chi^<_{\pm}&\approx&C\left( \begin{array}{c} 
  \re^{qW} \re^{i\pi W} \\
  0\end{array}\right)
\approx C\re^{qW}\left( \begin{array}{c} 
  \re^{i\pi W} \\
  0\end{array}\right)\label{xrm}
\eea

For $W$ odd ($N$ even) any two conduction bands $\chi^>_+,~\chi^<_+$ have the same 
asymptotic wave function (notice the ratio of $u_A/u_B=-1$). 
The same results holds for any two valence bands wavefunctions with a ratio $u_A/u_B=1$.
This means that as $k_x$ moves continuously through the band-center, the conduction band does not
cross the zero-energy (Fermi energy) point (the same holds for the valence band).
In contrast, for ribbons of $W$ even ($N$ odd), two  bands $\chi^>_+,~\chi^<_-$ with different energies are the
ones that have the same asymptotic wavefunction (notice the ratio of $u_A/u_B=-1$). Similarly, the
two other  bands $\chi^>_-,~\chi^<_+$ have their wavefunctions with the ratio $u_A/u_B=1$ as the center of the
band is approached. This means that if we change $k_x$ from left to right we go from conducting band to
valence band or from valence band to conducting band, i.e., the bands cross.

From Eqs.~(\ref{edge_p}) and (\ref{edge_m2}), we see that for odd $N$ the overlap between two spinors in the
conduction band at $k_{x}a = \pi \pm k$ is equal to zero: $\int dy <\chi_{+}(k_{x}a =\pi +k)|\chi_{+}(k_{x}a=\pi -k)> = 0$. On the contrary, for even $N$ we have $\int dy <\chi_{+}(k_{x}a =\pi +k)|\chi_{-}(k_{x}a=\pi -k)> = 0$.

The analysis of finite width ribbons presented here clearly shows that the origin of the even-odd dependence found in previous works in the limit of semi-infinite ribbons, is given by the nature of the band of the zero energy mode.

\section{Graphene ribbons as coupled quantum chains}
\label{sec2}

The even-odd dependence formalized in the previous section suggests that some physical properties of ZGR
should be captured by highly anisotropic models where a ribbon of finite width is viewed as a set of coupled
one-dimensional chains. As shown in Ref.\cite{ZS4}, the Hamiltonian of  $N$ decoupled ($t_{\perp}=0$) chains is  
$H_N=\sum_{n=1}^{n=N}H_n$ where $H_n=\int dx [t\bm c_{A_n}{\dag}(x)(\bm c_{B_n}(x)+\bm c_{B_{n}}(x-a))
+h.c]$ is the Hamiltonian of the $n^{th}$ chain. The energy of each chain is given by $E=2t\cos{k_xa/2}$. Near the center
of the band $k_{x}a = \pi$, we can write the lattice creation and annihilation operators by:
\bea
&& \bm c_{A_n}(x)=i(-1)^{x/a}(R_n(x)+L_n(x))/\sqrt{2}
\nn&&  \bm c_{B_n}(x)=(-1)^{x/a}(R_n(x)-L_n(x))/\sqrt{2}.
\label{Fradkin}
\eea

By introducing the Majorana fermions
\bea
&& R_n=(\xi_{2n-1}+i\xi_{2n})/\sqrt{2}
\nn && L_n=(\bar\xi_{2n-1}+i\bar\xi_{2n})/\sqrt{2}
\eea
the Hamiltonian becomes $ H_N=\int dx [{\cal H}_{0}+{\cal H'}] $
with
\bea
{\cal H}_{0}&=&\sum_{n=1}^{2N}iv(\xi_n\p_x\xi_n-\bar\xi_n\p_x\bar\xi_n), \label{H0N}\\
{ \cal H}' &=&\sum_{n=1}^{2N-2}- ih(\xi_{n}\xi_{n+2}-\bar\xi_{n}\bar\xi_{n+2})\\ 
&+& im(\xi_{n}\bar\xi_{n+2}-\bar\xi_{n}\xi_{n+2})
\label{Htp}
\eea
In graphene $h=m=t_{\perp}/2=t/2$. The spectrum of the Hamiltonian can be 
derived easily and matches well with exact
tight-binding results \cite{ZS4}.

We emphasize here that the origin of the mass term in ($\ref{Htp}$)  is due to the fact that in zigzag ribbons an $A$ atom from each
chain $n$ is connected to a $B$ atom in the next chain $n+1$
 but the $B$ atom in chain $n$ is not connected to an $A$ atom in the next
chain. In the
case of armchair ribbons, on the other hand, the connection between $A$
and $B$ atoms of adjacent chains is symmetric. However this argument
 does not exclude the
presence of the mass term in the effective Hamiltonian of armchair ribbons
 because  the spinor wavefunction of these ribbons has four components and
 contains two $A$ and two $B$ atoms.

\section{The square lattice model}
\label{sec3}

The generalized form of Eq.~(\ref{Htp}) with $h \neq m$, represents the square lattice
model in Fig.~(\ref{ZGR}) which has extra hopping terms between $B$ atoms
in any given chain and $A$ atoms in the next chain. The $m$ (mass) term represents the difference between
successive rung hopping terms,  $m=t_{\perp}(1-\eta)/2$, and the magnetic term $h$
represents the average $h=t_{\perp}(1+\eta)/2$.  Thus, the Hamiltonian in Eq.~(\ref{Htp}) describes a continuous
family of models that include the standard square lattice when $\eta = 1$ and the $\pi$-flux model when $\eta = -1$ \cite{Ludwig,Marston,Lieb}. For the particular case of graphene, values of $\eta \neq 0$ can represent various 
hopping terms. In particular the NNNN hopping term that has been estimated to be of the order of $t_3=0.1-0.3$eV \cite{White,Gonzalez}, can be incorporated in the renormalization of the
mass and magnetic terms producing $m=(t_{\perp}-3t_3)/2$ 
and $h=(t_{\perp}-t_3)/2$ respectively. Notice also, that this hopping mechanism is expected to be highly affected by
local lattice distortions produced either by ripples or bending of the material.

For the generalized case with $h \neq m$, following the steps outlined in Sec.~\ref{sec1}, we can solve the tight binding Hamiltonian and its wavefunction in a straightforward manner:
\be
H  =  \left( \begin{array}{cccc}
 0 & \varphi' \\
 \bar\varphi' & 0 
\end{array}\right)
~~~%
\Psi_{\pm}=C\left( \begin{array}{l}
u_A= \re^{i\alpha'/2}  \\
 u_B=\pm \re^{-i\alpha'/2}\end{array}\right)\re^{ik_xx+ik_yy}
\ee
with $\varphi'(k_x,k_y) =  t_{\perp}(\re^{ik_yb}+\eta\re^{ik_yb})+2t\cos{\frac{k_xa}{2}}$ and
$\bar\varphi'(k_x,k_y)=\varphi'(k_x,-k_y)$.
The angle $\alpha'$ is defined 
by $\varphi'(k_x,k_y)=\sqrt{\varphi'\bar\varphi'}\re^{i\alpha'}$.  
The energy of the model is given by
$E=\pm \sqrt{(t\cos{k_xa\over2}+2h\cos{k_yb})^2+4m^2\sin^2{k_yb} }$.
As before $k_x$ is a good quantum number  while  $k_y$ depends on $k_{x}$ and is determined 
by the boundary conditions, thus it is a function of $W$ and $k_x$.
In the presence of the $\eta t_{\perp}$ hopping term the boundary conditions have to be 
generalized so that both components of the spinor wave-function are zero at the edges. 
The wave-function inside the ribbon is constructed from the superposition of
four degenerate states at $k_y=(\pm k_1,\pm k_2)$:
\bea
&&\Psi_{ZGR}(k_x)=\sum_{k_y=\pm k_1,\pm k_2}a(k_y)\left( \begin{array}{l}
\re^{i\alpha'/2}  \\
\re^{-i\alpha'/2}\end{array}\right)\re^{ik_xx+ik_yy}
\nn&&
=C\left( \begin{array}{l}
\varphi(y)\nn
-\varphi(-y)
\end{array}\right)\re^{ik_xx}
\eea

After applying the boundary conditions we get
\bea
&&\varphi(y)=\sin\left({{k_2W-\alpha_2+\delta\over2}}\right)
\sin\left({2k_1y+\alpha_1+\delta\over2}\right)
\nn&-&\sin\left({{k_1W-\alpha_1+\delta\over2}}\right)
\sin\left({2k_2y+\alpha_2+\delta\over2}\right) 
\eea
with $\delta=0,\pi$. Here $k_1$ and $k_2$ are calculated by
the degeneracy condition $E(k_1)=E(k_2)$ and 
\bea
&&\sin\left({{k_1W-\alpha_1+\delta\over2}}\right)\sin\left({{k_2W+\alpha_2+\delta\over2}}\right)=
\nn&&\sin\left({{k_2W-\alpha_2+\delta\over2}}\right)\sin\left({{k_1W+\alpha_1+\delta\over2}}\right)
\eea
These two conditions determine $k_1$ and $k_2$ as functions of
$k_x$ and $W$. In general $k_1$ and $k_2$ are complex numbers. 

\begin{figure}[!]
 \includegraphics[width=.5\textwidth]{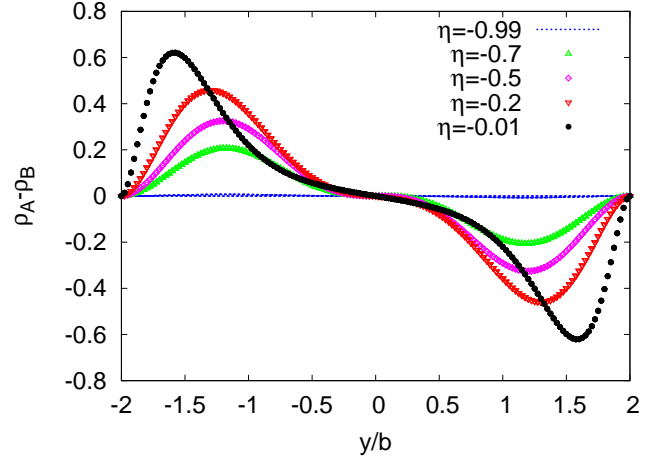}
 \caption{The plot of $\rho_A-\rho_B$ as a function of
$y$ across the ribbon with $W=4b$. As approaching the
$\eta=0$ point (graphene ribbon) the wave-function becomes
more localized.
}
 \label{rab}
\end{figure}

As mentioned above, the limit  $\eta=-1$  describes the $\pi-$flux phase that in this language implies the case 
with $h=0$. In this regime, the wavenumbers $k_1$ and $k_2$ are independent of $k_x$ and
are given by $k_1b=\pi/2+n\pi b/W$ and $k_2b=\pi-k_1b$. The energy takes the simple form:
$E=\pm \sqrt{t^2\cos{k_xa\over2}^2+t_{\perp}^2\sin^2{k_yb} }$. 
The valence and conduction bands coincide at four points of the Brilliouin
zone $(k_xa=\pm\pi,k_yb=\pm\pi)$.  At each point there are 
two  gapless modes with $E=\pm 2t\cos{k_xa\over2}\approx \pm tk$. 
Note that for  odd  $N$  one of the solutions of $k_y$ lies exactly at the gapless
corner of Brillouin zone. On the contrary
for even $N$ the zero energy points are always avoided. 
By increasing $\eta$, the wave numbers $k_1$ and $k_2$ become complex quantities 
and acquire a dependence on $k_x$. However, for odd $N$, the zero energy mode
survives  all along the line $-1\leq\eta\leq1$ with a dispersion that changes from
linear at $\eta=-1$ to Eq.~(\ref{EEd}) at  $\eta=0$. As Fig.~(\ref{rab}) shows, the wave functions of different modes
shows incipient localization as $\eta$ increases and become localized when  $\eta\to 0$. 
In the regime with $\eta>0$, the magnetic term dominates and introduces new zero-energy
points for both values of $N$, odd or even. However, for odd $N$, there is an odd number of
zero-energy points, one of them being the existing mode at $k_xa=\pi$. In contrast, for even $N$, there is an even 
number of zero-energy points with no zero modes at $k_xa=\pi$. Note that 
the extreme limit  $\eta\to1$ represents the simple square lattice
where  $k_yb=n\pi b/W$  and energy spectrum 
$E=\pm (t\cos{k_xa\over2}+t_{\perp}\cos{k_yb})$. In this case, the mode with linear dispersion at $k_xa=\pi$ exists only 
for ribbons with even $W/b$ (or odd $N$). 

In order to visualize the evolution of the band structures and wavefunctions of all  these different models, 
it is instructive to study the simplest examples of $ZGR$s
with $N=2$ (even) and $N=3$ (odd) number of chains. 

The Hamiltonian of $N=2$ ZGR   consists of two independent 
sets of Majorana pairs, odd and even: $(\xi_1,\xi_3)$ and $(\xi_2,\xi_4)$. Consider only
one of these sectors:
\bea
&&{\cal H}={\cal H}_0-ih(\xi_1\xi_3-\bar\xi_1\bar\xi_3)
\nn &&+im  (\xi_1\bar\xi_3-\bar\xi_1\xi_3)
\eea

By the canonical transformation
\be
\xi_1=(\chi_1+\chi_2)/\sqrt{2}~~\xi_3=(\chi_1-\chi_2)/\sqrt{2}
\ee
the above Hamiltonian transforms to
\bea
&&{\cal H}={\cal H}_0+im(\chi_1\bar\chi_1-\chi_2\bar\chi_2)
\nn &&+ih (\chi_1\chi_2-\bar\chi_1\bar\chi_2)
\eea
At $\eta=-1 (h=0)$ the spectrum consists of two degenerate  massive fermions with $E=\pm\sqrt{m^2+k^2}$.
For generic $\eta$  the magnetic term $h$ lifts the degeneracy and 
$E=\pm(\sqrt{m^2+k^2}\pm h)$.  Graphene ribbons are represented by the
critical value $h=m$ with the particular dispersion of edge states given by:
$E\approx k^2/2m$. For $\eta>0$ the bands cut the Fermi energy at two points
around $k_xa=\pi$.

For   ZGRs with $N=3$ the Hamiltonian (\ref{Htp})  
consists of two independent sets of three Majorana fermions (even and odd). 
Consider one of these sectors:
\bea
&&{\cal H}={\cal H}_0
+im  (\xi_1\bar\xi_3+\xi_3\bar\xi_5-\bar\xi_1\xi_3-\bar\xi_3\xi_5)
\nn &&-ih(\xi_1\xi_3+\xi_3\xi_5-\bar\xi_1\bar\xi_3-\bar\xi_1\bar\xi_5)
\eea
It is useful to define linear combinations of the Majorana fields $\xi_{1}$ and $\xi_{5}$ as:
$\eta_1=(\xi_1+\xi_5)/\sqrt{2}$ and $\eta_3=(\xi_1-\xi_5)/\sqrt{2}$. Working on the basis $\eta_{1}, 
\eta_{3}$ and $\xi_{3}$ it is easy to see that the mass term does not couple   $\eta_3$ to the   
two other modes $\eta_{1}$ and $\xi_3$ and thus, this mode remains massless.
By applying a second transformation:
\bea
&&\chi_1=(\xi_3+\eta_1)/\sqrt{2}~~~\bar\chi_1=(\bar\xi_3+\bar\eta_1)/\sqrt{2}
\nn&&\chi_2=(\xi_3-\eta_1)/\sqrt{2}~~~\bar\chi_2=(\bar\xi_3-\bar\eta_1)/\sqrt{2}
\eea
The Hamiltonian transforms into:
\be
{\cal H}={\cal H}_0+im\sqrt{2}(\chi_1\bar\chi_1-\chi_2\bar\chi_2)
+ih(\chi_1\eta_3+\chi_2\eta_3-\bar\chi_1\bar\eta_3-\bar\chi_2\bar\eta_3)\label{HS3}
\ee
At $\eta=-1 (h=0)$ the spectrum has two linear modes $E=\pm k$ 
and two copies of  massive modes with dispersion
$E=\pm \sqrt{k^2+2m^2}$. The linear modes correspond to the $k_yb=\pi$ modes
of the $\pi-$ flux lattice model. An increase of $\eta$, introduces the magnetic term that 
couples these modes to  the massive modes. As a result, the dispersion of the
massless modes changes, however they remain massless.
 
Finally, the  energy spectrum of the model for a generic value of $\eta$ is given by
\be
E(E^2-k^2-2m^2-2h^2)=\pm k(E^2-k^2-2m^2+2h^2).
\ee
When $h=m$ (at $\eta=0$) the massless modes dispersion becomes 
 $E\sim \pm k^3/4m^2$ which describe the edge state of the corresponding $ZGR$ ribbon. In the regime $\eta>1$ 
 two extra zero-energy points are introduced. As $\eta\to 0$ the
dispersion of the massless modes becomes linear again. These linear modes
are in fact the solution of the square lattice spectrum near the corners of the Brilliouin 
zone $(k_xa=\pm\pi,k_yb=\pm\pi/2)$.

\section{Examples of single-particle interaction terms}
\label{sec4}

Another useful aspect of the coupled chains representation is that it allows to study the effect of different single-particle
terms in the Hamiltonian and to obtain the corresponding band-structures in a rather straightforward manner.
To exemplify these points, in this section we derive the Majorana representation of three such terms: chemical potential, 
second neighbor hopping term and spin-orbit interactions. We also present the corresponding expressions for terms that 
involve spin-orbit interactions, focusing on the intrinsic and Rashba spin-orbit terms.

It is interesting to notice that the mapping to Majorana fermions, preserves the
quasi-degenerate features of the edge bands while
captures the main effects of these terms in the model for $ZGR$s. At the same time, the total 
Hamiltonian becomes richer and, as we discuss in the next section, some of
these terms have direct interpretations in terms of models of quantum spin chains.

From Eq.~(\ref{Fradkin}) we can obtain the relation between the fermion lattice operators $c_{A/B}$ and
the Majorana fermions representation as: 
\bea
c_{A_n}(x)&=&(-1)^x\frac{[i(\xi_{2n-1}+\bar\xi_{2n-1})-(\xi_{2n}+\bar\xi_{2n})]}{2}
\nn c^{\dag}_{A_n}(x)&=&(-1)^x\frac{[-i(\xi_{2n-1}+\bar\xi_{2n-1})
-(\xi_{2n}+\bar\xi_{2n})]}{2}
\nn c_{B_n}(x)&=&(-1)^x\frac{[(\xi_{2n-1}-\bar\xi_{2n-1})+i(\xi_{2n}-\bar\xi_{2n})]}{2}
\nn c^{\dag}_{B_n}(x)&=&(-1)^x\frac{[(\xi_{2n-1}-\bar\xi_{2n-1})-i(\xi_{2n}-\bar\xi_{2n})]}{2}
\eea

 The reverse relations read:
 \bea
 && \xi_{2n-1}+\bar\xi_{2n-1}=i(-1)^x(c^{\dag}_{A_n}(x)-c_{A_n}(x))
 \nn&&\xi_{2n}+\bar\xi_{2n}=-(-1)^x(c^{\dag}_{A_n}(x)+c_{A_n}(x))
 \nn&&\xi_{2n-1}-\bar\xi_{2n-1}=(-1)^x(c^{\dag}_{B_n}(x)+c_{B_n}(x))
 \nn&&\xi_{2n}-\bar\xi_{2n}=i(-1)^x(c^{\dag}_{B_n}(x)-c_{B_n}(x))
 \eea

A chemical potential term is given by ${\cal H}_{\mu}(x)=\mu\rho_+(x)=\mu(\rho_{A_n}(x)+\rho_{B_n}(x))$. In the Majorana fermion language it reads: 
\bea
&&{\cal H}_{\mu}\approx i2\mu(\xi_{2n-1}\xi_{2n}+\bar\xi_{2n-1}\bar\xi_{2n})
\eea
The effect of this term is trivial: it shifts the Fermi energy from $E_F=0$ to $E_F=\mu$ as expected.

\begin{figure}[!]
 \includegraphics[width=0.5\textwidth]{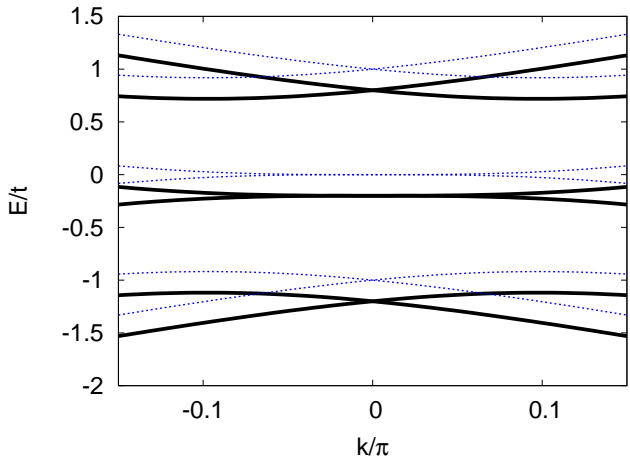}
 \caption{Energy bands of the zigzag ribbon with  $N=3$
   in the presence (solid lines) and absence (dotted lines) of 
   the next nearest hoping term $t_2=0.1t$.
 }\label{t2}
\end{figure}

A second-neighbor hopping (NN) term is given by:
${\cal H}_{NN}=t_{2} \sum_{<ij>} (c^{\dag}_{A_i}(x) c_{A_j}(x) + c^{\dag}_{B_i}(x) c_{B_j}(x)) + h.c. $. The coupling constant 
$t_{2}$ has been estimated to take its value in the range $t_2=0.02 t - 0.2 t$ eV\cite{Neto}. In terms of Majorana fermions, this term introduces a velocity renormalization (which can be ignored) due to the inter-chain coupling and an intra-chain contribution given by: 
\bea
&{\cal H}_{NN} \approx -i2t_2(\xi_{2n-1}\xi_{2n}+\bar\xi_{2n-1}\bar\xi_{1n}).
\eea
This expression suggests that the NN hopping term, in a first approximation, acts as an effective chemical potential, thus breaking particle-hole symmetry \cite{Saito,Sasaki}. The edge state energy is lowered with the corresponding modes 
acquiring a finite velocity along the ribbon and becoming more stable Fig.(\ref{t2}).

\begin{figure}[!]
 \includegraphics[width=0.5\textwidth]{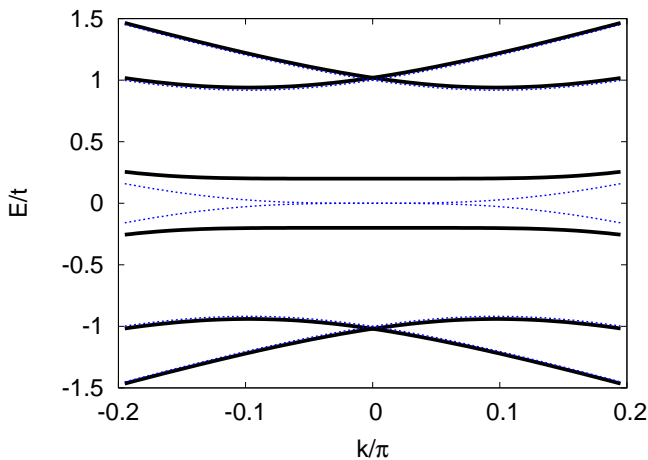}
 \caption{Energy bands of the zigzag ribbon with  $N=3$
   in the presence (solid lines) and absence (dotted lines) of 
   the next staggered sublattice potential term $\nu=0.2t$.
 }\label{nu}
\end{figure}

A staggered sublattice potential is  represented by a Hamiltonian ${\cal H}_{\nu}(x)=\nu\rho_-(x)=\nu(\rho_{A_n}(x)-\rho_{B_n}(x))$. In the Majorana fermion representation it takes the form:
\bea
&&{\cal H}_{\nu}(x)
\approx i2\nu(\xi_{2n-1}\bar\xi_{2n}+\bar\xi_{2n-1}\xi_{2n})
\eea
For the ZGR in terms of coupled chains, this term opens a gap in each single chain (even when decoupled from the rest, with
$t_{\perp}=0$). Thus, every chain making the ribbon is
gapped or equivalently,  each pair of Majorana fermions is massive. The 
total spectrum of the ribbon becomes massive regardless of  $N$ being odd or 
even (\ref{nu}). Note that this term corresponds to a large momentum transfer between two in-equivalent Dirac points.

Notice that, as a consequence, in $N$-odd ribbons, the only term which can open a gap at $k_{x}a = \pi$, involves 
inter-valley scattering. Although in graphene the staggered chemical potential is expected to be zero, recent experimental work on suspended graphene samples suggests the presence of a small energy difference between the two sub-lattices\cite{Eva}.
Moreover in nanoribbons made of  $SiC$  or  $BN$ there is an intrinsic staggered chemical potential due to the presence of two different atoms in sublattices $A$ and $B$. \cite{Guo,Zheng,Sun}

\begin{figure}[!]
 \includegraphics[width=0.5\textwidth]{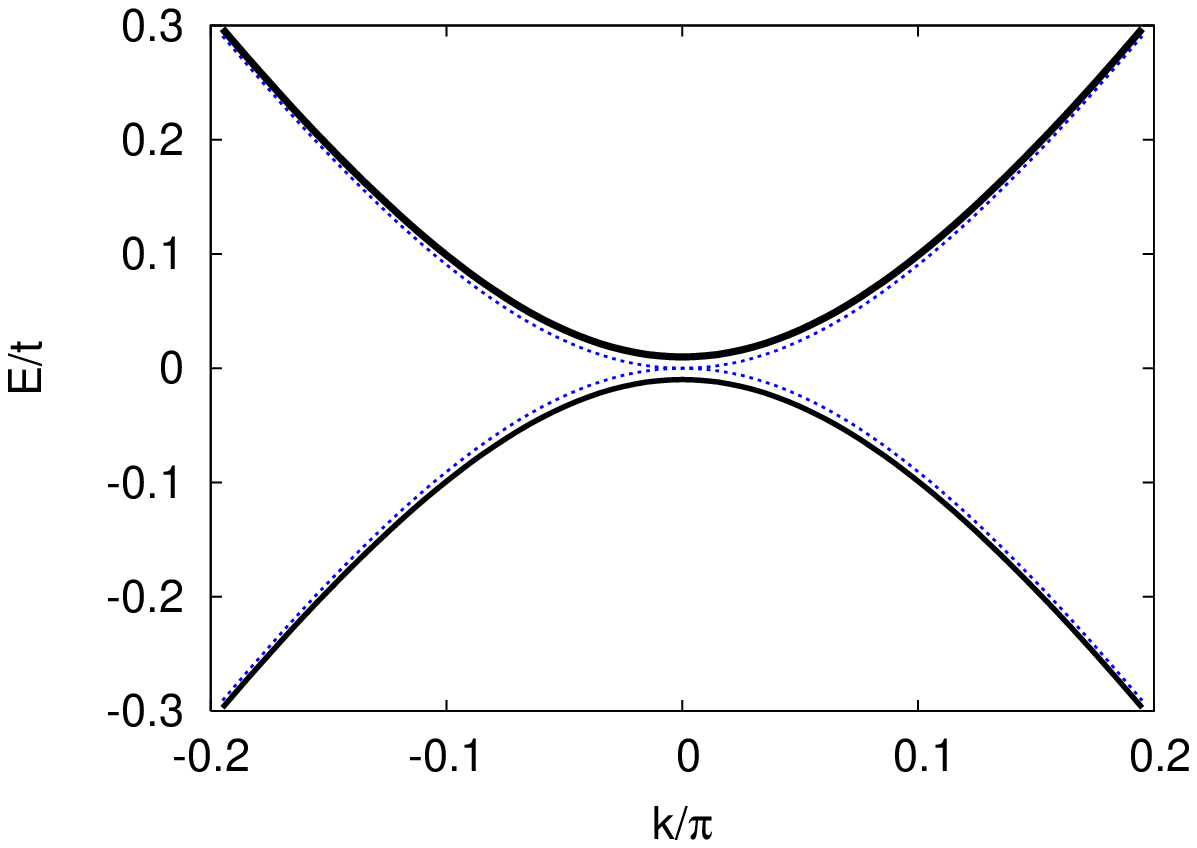}   
\includegraphics[width=0.5\textwidth]{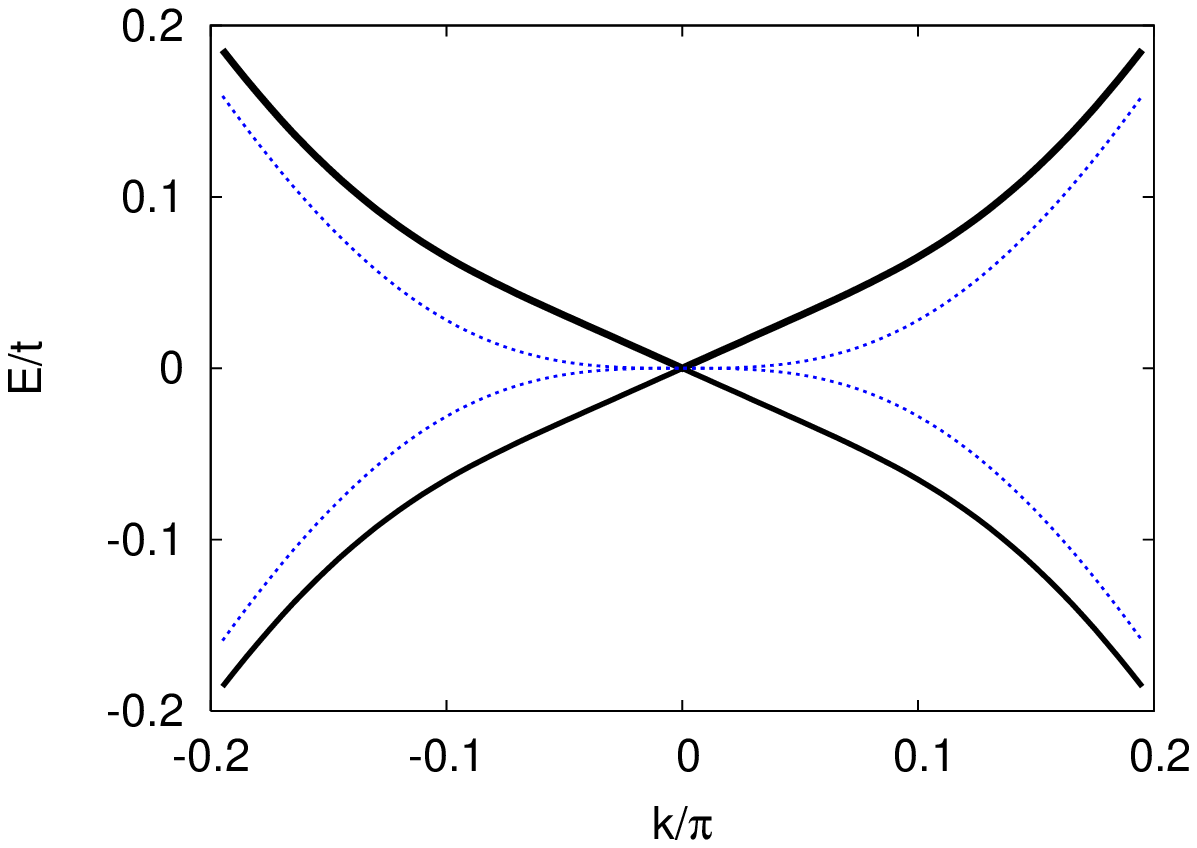}
 \caption{Energy bands of the zigzag ribbon with  $N=2$ (top) and$N=3$ 
(bottom)
   in the presence of
   the intrinsic spin-orbit  $t'=0.1t$. For even N there is a gap at 
$k_xa=\pi$.
 }\label{g}
\end{figure}

The procedure can also be applied to other single-particle terms in the Hamiltonian. For example, 
we can also study the expression of spin-orbit interaction terms using the Majorana representation. 
The intrinsic spin-orbit (I-SO) is $H_{ISO}=it's\sum_{<ij>}v_{ij}c_{i}^{\dag}c_j$ in
which $s=\pm$ refers to the z-component of the electron spin operator.
$v_{ij}=2(d_{ik}\times d_{kj})/\sqrt{3}$ is the path-dependent number which takes the values $v_{ij}=(\pm 1)$
depending on the shortest path for a hopping process that goes from site $i$ to the next 
nearest cite $j$ through an intermediate cite $k$\cite{Kanemele,Kanemele2,ZS1}. 
For modes with spin up the spin-orbit is written as 
\bea
&&H_{ISO}\approx t'(\xi_n\bar\xi_{n+2}+\bar\xi_{n}\xi_{n+2}).\label{HISO}
\eea
For spin down the equivalent term corresponds to the transformation $t'\to-t'$.  It has been shown that in the presence of the I-SO interaction, the bulk states have a gap of the order of $t'$. The situation is different in ribbons. For ZGRs with $N$ odd, the edge states get exchanged as the conduction and valence bands cross each other at the midgap point $k_xa=\pi$. The ribbon thus remains metallic. 
In this case the I-SO interaction removes the degeneracy of the edge states near $k_xa=\pi$ and  renders a pair of linear dispersion bands. Moreover, the edge states become 
spin-filtered with opposite spin currents at opposite edges of the ribbon. This is the topological insulating or quantum spin Hall phase of graphene  \cite{Kanemele}. For ZGRs with even $N$ the ISO  interaction opens a gap between the edge states at $k_xa=\pi$ \cite{ZS2}. The gap is very small and scales as $\Delta\sim (t'/t)^N$. Graphene ribbons in this phase are normal insulators Fig.~(\ref{g}).

For graphene samples on a substrate or suspended but subjected to a
perpendicular electric field, the Rashba spin-orbit (RSO) interaction becomes quite relevant. 
The expression for the RSO interaction is $H_{RSO}=i\lambda \sum u_{ij}c_i^{\dag}c_j$
where $u_{ij}\sim d_{ij}\times {\cal E}_z$. Here $d_{ij}$ is a vector which
connect the nearest neighbor cites $i$ to $j$. The RSO has two contributions:
the intra-chain part gives
\bea
&&H_{RSO1}=-i4\sqrt{3}\lambda(\xi_{2n\up}\xi_{2n\down}-\bar\xi_{2n\up}\bar\xi_{2n\down}
\nn&&+\xi_{2n-1\up}\xi_{2n-1\down}-\bar\xi_{2n-1\up}\bar\xi_{2n-1\down})
\eea
The inter-chain part is
\bea
&&H_{RSO2}=i2\lambda(\xi_{2n-1\up}\xi_{2n+2\down}-\bar\xi_{2n-1\up}\bar\xi_{2n+2\down}
\nn&&+\xi_{2n-1\down}\xi_{2n+1\up}-\bar\xi_{2n-1\down}\bar\xi_{2n+1\up}
\nn&&-\xi_{2n\up}\xi_{2n+1\down}+\bar\xi_{2n\up}\bar\xi_{2n+1\down}
\nn&&-\xi_{2n\down}\xi_{2n+1\up}+\bar\xi_{2n\down}\bar\xi_{2n+1\up}
\nn&&-\xi_{2n-1\up}\bar\xi_{2n+2\down}+\bar\xi_{2n-1\up}\xi_{2n+2\down}
\nn&&-\xi_{2n-1\down}\bar\xi_{2n+2\up}+\bar\xi_{2n-1\down}\xi_{2n+2\up}
\nn&&+\xi_{2n\up}\bar\xi_{2n+1\down}-\bar\xi_{2n\up}\xi_{2n+1\down}
\nn&&+\xi_{2n\down}\bar\xi_{2n+1\up}-\bar\xi_{2n\down}\xi_{2n+1\up})
\eea

\begin{figure}[!]
 \includegraphics[width=0.5\textwidth]{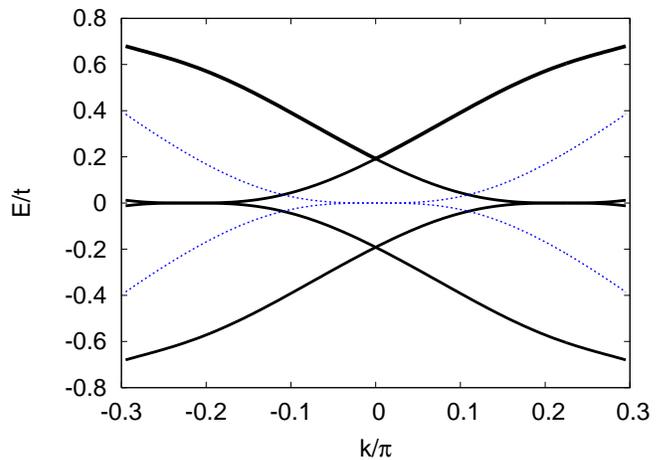}
 \caption{Energy bands of the zigzag ribbon with  $N=3$
   in the presence (solid lines) and absence (dotted lines) of 
   the Rashba spin orbit interaction $\lambda=0.1t$.
 }\label{tr}
\end{figure}

Notice that the RSO does not open a gap in the bulk\cite{ZS3}. The same holds for ZGRs: there is no gap in the ribbon' spectrum and the dispersion remains power-law like near the zero-energy point. However, in the presence of RSO interactions,   the zero-energy point located at $k_xa=\pi$ for RSO zero, shifts in opposite directions for opposite spins\cite{ZS3} Fig.~(\ref{tr}). For wider ribbons with quasi-flat (and originally spin-degenerate) bands, there is a more pronounced lift of the spin-degeneracy.
Another interesting point is that the RSO interaction does not remove the even-odd width dependence of the edge states bands for finite widht ZGRs. As a consequence even in the presence of RSO a ribbon with odd number of chains remains metallic while one with even number of chains is insulating. Notice also that these results have important consequences for the analysis of the topological insulator phases that can be observed in graphene ribbons as discussed in Ref.~\cite{Kanemele2}. This can be readily seen by considering an even $N$ ribbon: in this case, there is an even number of Kramer pairs at each edge. In contrast, for odd $N$ ribbons, the number of Kramer pairs is always odd, even in the presence of RSO interactions.
  
\section{Graphene ribbons and quantum spin chains models}

The Hamiltonian (\ref{Htp}) can also describe  some specific quantum spin chains models.
Before presenting the connection between the two approaches, it is instructive to briefly review  
some results of the models of: two coupled spin-$1/2$ chains and a spin-1 chain. \cite{bBook}

a) {\it Two coupled spin-$1/2$ chains.}

The symmetry of two spin-1/2 Heisenberg antiferromagnetic chains, in the absence of the 
inter-chain coupling, is $SU(2)\times SU(2)$. In the continuum limit the
spin operator in each chain, can be written in terms of current operators and a staggered part as $\bm S_i(x)=\bm J_i(x)+\bm{\bar J_i}(x)+(-1)^{x/a}\bm n_i(x)$ where $i=1,2$ labels chain $i$.
The smooth part of the spin operator corresponds to the sum of right and left mover
currents ($\bm J+ \bar{\bm J}$) of the $SU(2)_{k=1}$ WZNW model, and represents
the total magnetization of the chain.The staggered magnetization is represented by  $\bm n$ and  generalized 
models contain a dimerization operator $\epsilon=(-1)^{x/a}\bm S(x).\bm S(x+a)$
that represents the time component of the staggered magnetization \cite{Nersesyan}.

The addition of an interchain coupling reduces the symmetry to
the generic $SU(2)\times Z_2$. In the standard model for two spin-$1/2$ coupled chains (ladder)
\cite{Shelton}, the interchain coupling is isotropic:
$H'=J_{\perp}\bm S_1.\bm S_2$. The relevant part of
this term  stems from the staggered magnetization contribution and is given by  
$H'\approx J_{\perp}\bm n_1.\bm n_2$. The consequence of including such a term is the opening of a gap in the spectrum. 
In terms of Majorana fermions, the total Hamiltonian can be written in terms of four fermion fields $\xi_1, \xi_2, \xi_3$ and $\xi_4$. Three of these fields $\vec \xi=(\xi_1,\xi_2,\xi_3$) have equal masses $m\sim J_{\perp}$
and the remaining one ($\xi_4$) has a mass $-3m$. The classification of Majorana fermion states into
triplet and singlet states reflects the underlying global $SU(2)$ symmetry of the ladder.  

Majorana  fermions $\vec\xi$ describe the triplet excitations  
of the two-coupled $S=1/2$ quantum chain \cite{Tsvelik}. When the model is written in terms of right and left
movers it can be shown that the total currents  $\bm I=\bm J_1+\bm J_2$  for each type of mover
satisfy the Kac-Moody algebra of the $SU_{k=2}(2)$  vector currents. 
Since the central charge of the model is $C=3/2$, the currents can be represented using a 
triplet of Majorana  fermions: $\bm I^a={-i\over2}\epsilon^{abc} \xi_b \xi_c$ 
where $a,b,c$ refers to $1,2,3$. The model has two primary fields 
namely $\Phi^{(1/2)}$ and $\Phi^{(1)}$, with the second field 
given by  $\Phi^{(1)}_{ab}=\xi_a\bar\xi_b$. The other Majorana field $\xi_4$ describes the singlet excitations of the model. 
This  field $\xi_4$ contributes to the relative (or the so called 
wrong\cite{bBook}) currents $\bm K=\bm J_1-\bm J_2$    
which are represented by $\bm K^a=i\xi_a\xi_4$. 
It is convenient at this point to introduce the fields $Q_a=\xi_4\bar\xi_a$ 
and  $\bar Q_a=\bar\xi_4\xi_a$. 

Notice that the ferro-antiferromagnetic transformation
$J_{\perp}\to-J_{\perp}$ (which is reflected in $m\to -m$) does not change 
the thermodynamics but it interchanges singlet and triplet roles. 

b) {\it Spin $S=1$ chain}

The connection to a $S=1$ Heisenberg spin chain is given by considering the 
biquadratic $S=1$ chain as defined by \cite{Solyom,Tsvelik-90}
\be
H=\sum_x[S(x).S(x+a)+b(S(x).S(x+1))^2]
\ee
At $b=-1$ the model has gapless modes and it is equivalent to a WZW $SU(2)_{k=2}$
model. The Hamiltonian at this point can be described by Majorana fermions as
\be
H_0=\sum_{n=1}^3iv(\xi_n\p_x\xi_n-\bar\xi_n\p_x\bar\xi_n)
\ee 
For $b<-1$ there are two degenerate massive dimerized phases. For $-1<b<1$ the
spin chain is in the Haldane phase with a gapped singlet ground state
\cite{Nijs,Kennedy}.  

The link between the spin operator and the WZW model is given by
\be
\vec S(x)\sim \bm I(x)+\bar{\bm I}(x) +(-1)^x \bm n(x) 
\ee
Here $\bm n$ is the staggered part of the spin field. In analogy with the coupled spin-$1/2$ chains, 
the smooth (average) magnetization is given by $\bm M=\bm I+\bar {\bm I}$. 

For $0<b<-1$ the model has a gap that increases monotonically as
a function of $m=-1-b$. Close to $b=-1$ the model can be described effectively by
adding relevant mass terms. The mass term is positive in the dimerized phase
and negative in the Haldane phase. These mass terms are expressed in terms of the relevant fields
$\Phi^1_{ab}=\xi_a\bar\xi_b$ with a $3\times3$ matrix representation.  
For isotropic models the mass term is written by $mTr(\Phi^1)$.

c) {\it ZGRs and quantum spin chains}

To show the connections between these models and graphene ribbons we start with the even $N=2$ ZGR. 

The Hamiltonian written in terms of right and left movers can be bosonized in a straightforward manner by  
introducing the bosonic fields $\phi_+ = \exp(\pm i\sqrt{4\pi}\phi_+)=(\xi_1+i\xi_3)/\sqrt{2}$
and  $\bar\phi_+=\exp(\pm i\sqrt{4\pi}\bar\phi_+)=(\bar\xi_1+i\bar\xi_3)/\sqrt{2}$ for one set of movers and the corresponding $\phi_-, \bar\phi_-$ for the other. The total Hamiltonian density  reads 
\be
{\cal H}={\cal H}_0+i{m\over\pi}\sum_{\nu=\pm}\sin\sqrt{4\pi}\Theta_{\nu}
+{h\over\sqrt{2\pi}}\sum_{\nu=\pm}\nu\p_x\Theta_{\nu}
\label{2zgr}
\ee
where  $\Theta_{\nu}=\phi_{\nu} -\bar\phi_{\nu}$ is the dual field of
$\Phi_{\nu}=\phi_{\nu} +\bar\phi_{\nu}$. 
Equation (\ref{2zgr})
describes two decoupled
sine-Gordon Hamiltonianas with a magnetic field (or chemical potential)
applied to the dual sector.  

The model for the $N=2$ ribbon can be mapped to the two-leg ladder Hamiltonian discussed above with an anisotropic
interchain coupling and an in-plane applied magnetic field:
${\cal H}={\cal H}_0+{\cal H}'_{1h}+{\cal H}'_{1m}+{\cal H}'_{2m}$.
The magnetic field term is expressed in terms of the current operators as
${\cal H}'_{1h}=h(\bm K^y-\bar{\bm K}^y+\bm I^y-\bar{\bm I}^y)$ that are written in terms of Majorana fermions as shown above. The operator content of these currents can be given in terms of vector 
spin-chirality operators as:
$\bm I-\bar{\bm I}\sim \bm S_1(x)\times\bm S_1(x+a)+ \bm S_2(x)\times\bm S_2(x+a) $ and 
$\bm K-\bar{\bm K}\sim \bm S_1(x)\times\bm S_1(x+a)- \bm S_2(x)\times\bm S_2(x+a)$. 
The mass terms are produced by the following two interchain couplings.
The first term is ${\cal H}'_{1m}=im(Q_{2}-\bar Q_{2})$. Expressed as 
spin operators this is  $Q_{2}-\bar Q_{2}\sim (\bm S_1\times\bm S_2)^y 
\sim (\bm n_1\times\bm n_2)^y$.
The second term is ${\cal H}'_{2m}=im(\Phi^{(1)}_{13}+\Phi^{(1)}_{31})$
which can be generated from
$(\Phi^{(1)}_{13}+\Phi^{(1)}_{31})\sim (n^z_1n^x_2+n^z_2n^x_1)$. 

The analysis can be extended to the next even ZGR ribbon, namely the $N=4$ ribbon
which can be described by two sets of four Majorana fermions. 
The Hamiltonian of each sector,  separately, can be represented 
by a two-chain spin-$1/2$ ladder. To get the equivalent spin chain model, we 
just need to add two more terms to the equivalent spin model for the $N=2$ ZGR discussed above: 
The first term is  ${\cal H}'_{3m}=im(Q_1-\bar Q_1)\sim
t_{\perp}(\bm S_1\times\bm S_2)^x$. The second term is
the magnetic field ${\cal H}'_{2h}=h(\bar{\bm K}^x-{\bm K}^x)$.

The mapping between odd ZGR ribbons and spin-chain models is simpler for the $N=3$ case.
The equivalent spin-chain describes the anisotropic version of the biquadratic spin $S=1$ chain\cite{Tsvelik-90} 
with an in-plane magnetic field. The Hamiltonian of one of the sectors is given in (\ref{HS3}).
This model describes a $S=1$ model with masses $m_1=m$, $m_2=-m$
and $m_3=0$ and two magnetic fields $h_1=h_2$. 
In the language of ZGR ribbons, we notice that the value of the mass terms $m_1$ and $m_2$ can be changed when
including the ISO interaction (\ref{HISO}). In this case then, 
the mass terms are given by $m_1=t_{\perp}/2+st'$ and $m_2=-t_{\perp}/2+st'$
where $s=\pm$ stands for real spin-up spin-down electron. 

For the $N=5$ ZGR  each sector has five Majorana fermions. In principle 
the Hamiltonian can be described by the $S=2$ representation of the
$SU_{k=10}(2)$ WZW model. The corresponding primary field has the conformal dimension 
$(1/2,1/2)$ which can be written in term of Majorana fermions bilinears. 
Furthermore, the central charge of the theory is $C=5/2$ which is another indication that five Majorana fermions are 
needed to describe the model. 

\section{Conclusions}

Zigzag graphene ribbons show remarkable physical characteristics that are predicted to have important effects in their transport properties. In this work we have provided a detailed analytic treatment of a tight-binding Hamiltonian with hard-wall boundary  conditions that explains the surprising width-dependent properties in terms of the existence of a zero-energy mode that corresponds to localized states along the edges of a finite width ribbon. As a consequence, ZGRs with even widths (odd number of chains) are metallic while odd width ribbons (even number of chains) are insulating. 

The unusual even-odd dependence led us to consider models of ZGR in terms of coupled quantum chains. This particular continuous limit preserves the main features of band structures and spinor wave-functions while keeping the width-dependence.

In the Majorana fermion representation, the model of ZGR in terms of coupled chains can be easily extended to include various single-particle interactions and we have analyzed in details the effects of adding a chemical potential term, second and third nearest neighbor  hoppings, a staggered chemical potential, intrinsic spin-orbit and Rashba spin-orbit interactions.

Besides the straight-forward treatment of these terms, the mapping reveals that ZGRs can be viewed as a member of a continuous family of models in square lattices that range from the standard square lattice to $\pi-$flux models. These models share the same even-odd width dependence first obtained in ZGR models. We have presented a full solution for these models and emphasized their similarities with ZGRs.

As a final application, we have shown that it is possible to treat ZGRs as coupled quantum spin chains. A careful analysis of models in terms of Majorana fermions shows that ZGR with two chains inside are fully equivalent to two spin-$1/2$
coupled chains with an anisotropic interchain coupling and an external magnetic field. We also analyzed another example of
an even-width ribbon ($N=4$). 
We applied the mapping to odd-number models and showed that an $N=3$ ribbon, corresponds to the anisotropic biquadratic spin $S=1$ chain with an in-plane magnetic field. A preliminary analysis of an $N=5$-chain ribbon suggests a description in terms of a set of 5 Majorana fields for each Hamiltonian sector.
We would like to remark that the mapping proposed here allows to obtain solutions for all these new quantum spin chains models by direct comparison with the known results obtained with ZGR's.

\section{Acknowledgements}
We acknowledge A. W. W. Ludwig,  F. Guinea and H. Johannesson for useful discussions. This work was partially supported by NSF under grants DMR-0710581 and PHY05-51164; and Ohio University BNNT funds.

\end{document}